\documentclass[useAMS,usenatbib]{mn2e}
\usepackage{graphicx}

\title[Cherenkov drift emission of pulsars] {Very high energy emission of Crab-like pulsars driven by the Cherenkov drift radiation.}

\author[Osmanov Z.]{Osmanov Z.\thanks{E-mail:
z.osmanov@freeuni.edu.ge}\\
School of Physics, Free University of Tbilisi, 0183, Tbilisi,
Georgia\\
}
\begin{document}

\pagerange{\pageref{firstpage}--\pageref{lastpage}} \pubyear{2011}

\maketitle

\label{firstpage}

\begin{abstract}
In this paper we study the generation of very high energy (VHE)
emission in Crab-like pulsars driven by means of the feedback of
Cherenkov drift waves on distribution of magnetospheric electrons.
We have found that the unstable Cherenkov drift modes lead to the
quasi-linear diffusion (QLD), keeping the pitch angles from
vanishing, which in turn, maintains the synchrotron mechanism.
Considering the Crab-like pulsars it has been shown that the growth
rate of the Cherenkov drift instability (ChDI) is quite high,
indicating high efficiency of the process. Analyzing the mechanism
for the typical parameters we have found that the Cherenkov drift
emission from the extreme UV to hard $X$-rays is strongly correlated
with the VHE synchrotron emission in the GeV band.
\end{abstract}


\section{Introduction }
Last decade results from the new facilities: Very Energetic
Radiation Imaging Telescope Array System (VERITAS) and Major
Atmospheric Gamma-ray Imaging Cherenkov (MAGIC) Telescope have
significantly stimulated an interest to VHE gamma ray observations.

In particular, the VERITAS collaboration has reported the pulsed
gamma ray emission from the Crab pulsar above $100$GeV
\citep{veritas}, which later have been confirmed by the MAGIC
observations \citep{magic1,magic2}. It is obvious that such
energetic photons are emitted by VHE particles. On the other hand,
the pulsar magnetosphere is mainly composed of electron-positron
pairs which, by means of the centrifugal mechanism of acceleration
might achieve extremely high energies in the light cylinder area (a
hypothetical zone, where the linear velocity of rotation exactly
equals the speed of light) \citep{or09,screp1,screp2}. Usually, the
major question which arises in the context of these observations is
to understand and identify concrete mechanisms responsible for
generation of VHE photons.

In general, it is strongly believed that VHE gamma-rays are produced
either by the inverse Compton or by the curvature radiation
mechanism. In particular, on the light cylinder surface in the
magnetospheres of the Crab-like pulsars the magnetic field is of the
order of $B\sim 10^6$G, which efficiently suppresses the synchrotron
mechanism of radiation. Indeed, the cooling timescale for this
process is given by $t_{syn}\sim\gamma mc^2/P_{syn}$, where $\gamma$
is the Lorentz factor of an electron, $m$ is the electron's mass,
$c$ is the speed of light, $P_{syn}\approx
2e^4\gamma^2B^2\sin^2\psi/3m^2c^3$ is the electron's synchrotron
emission power, $e$ is the electron's charge and $\psi$ is the pitch
angle. The above expression for $t_{syn}$ applied to the
relativistic electrons with $\gamma\sim 10^7$ \citep{or09} for
$\sin\psi\sim 1$ leads to the cooling timescale of the order of
$10^{-10}$s, which is by many orders of magnitude less than the
kinematic timescale, or the escape timescale $t_{kin}\sim
R_{lc}/c\sim P_{_{Cr}}/2\pi$, where $R_{lc}$ is the light cylinder
radius and $P_{_{Cr}}\sim 0.033$s is the Crab pulsar's rotation
period. This means that particles very rapidly transit to the ground
Landau states and as a result the emission process should damp.

In a series of papers \citep{mo09,mo10,osm11,cmo11,co12,cmo13} the
authors have shown that the pulsar magnetospheres might provide
necessary conditions for keeping the pitch angles from damping,
thus, maintaining the synchrotron mechanism. The major process
responsible for keeping the synchrotron emission from damping is the
QLD. In particular, \cite{kmm} have shown that plasmas with strong
magnetic field may generate very unstable cyclotron waves, which in
turn, via the QLD influence the particle distribution across the
magnetic field lines \citep{lomin}. This process inevitably leads to
the creation/maintaining of the pitch angles making the synchrotron
radiation a working mechanism despite the efficient energy losses.
Another interesting feature of this mechanism is that the QLD
provides emission in two different frequency bands: one generated by
the cyclotron modes and another produced directly by the synchrotron
radiation.

In general, distribution of particles might be influenced not only
by the cyclotron waves but also by the Cherenkov-drift waves,
finally driving the synchrotron process. In particular, in our
recent work \citep{oc13} we have considered the synchrotron
emission, generated by means of the feedback of Cherenkov drift
waves on the particle distribution in the active galactic nuclei. It
has been argued that in the light cylinder area the Cherenkov drift
instability is strong enough to prevent the pitch angles from
damping, keeping the synchrotron mechanism active and providing
strongly correlated emission in different energy bands.

In the present paper we consider the Crab-like pulsars and study the
synchrotron process maintained by means of the QLD driven by the
Cherenkov drift instability.

The paper is arranged in the following way. In section~II, we
introduce the theory of the Cherenkov driven synchrotron emission.
In section~III, we apply the model to Crab-like millisecond pulsars
and obtain results and in section~IV, we summarize them.

\begin{figure}
  \centering {\includegraphics[width=7cm]{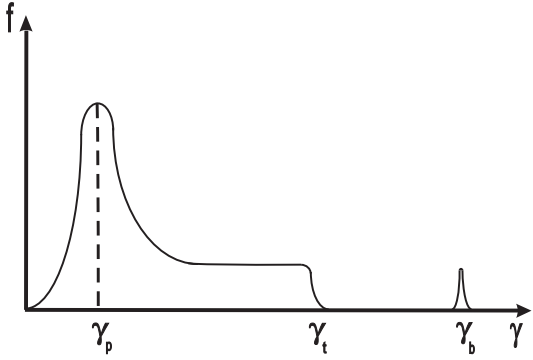}}
  \caption{Magnetospheric electrons' distribution function versus the Lorentz factor.
  The function consists of two major sections: the first -
  wider area corresponds to the so-called plasma component and the relatively narrower
  region corresponds to the primary Goldreich-Julian beam particles.}\label{fig0}
\end{figure}
\section[]{Emission model}

In this section we consider the light cylinder area of
magnetospheres of the Crab-like pulsars and develop the model of
synchrotron emission maintained by means of the feedback of the
Cherenkov drift instability. It is strongly believed that in the
mentioned zone by means of a direct centrifugal mechanism of
acceleration electrons might achieve Lorentz factors of the order of
$10^7$ \citep{or09}. In the framework of the paper we suppose that
the magnetospheric particles are distributed according to the Fig.
\ref{fig0}. The distribution function is composed of two major
parts. The narrower "region" represents the primary beam components
and the wider "area" is a result of the pair cascading processes,
which is considered in detail by \cite{stur,tadem}. By $\gamma_b$ we
indicate the Lorentz factor of the beam component, $\gamma_t$ is the
Lorentz factor of the tail and $\gamma_p$ represents the Lorentz
factor in the wider "section" of the plot.

Particles moving in a medium with curved magnetic field lines
experience drift across the plane of the curvature and the
corresponding velocity is given by
\begin{equation}\label{ud}
u_x = \frac{\gamma_bc^2}{\omega_B\rho},
\end{equation}
where $\omega_B = eB/(mc)$ represents the cyclotron frequency and
$\rho$ is the curvature radius of the field line. From the above
equation it is clear that for ultra relativistic electrons the drift
velocity might become relativistic as well and as is shown by
\cite{shapo1}, under these circumstances the ChDI arises with the
following resonance condition
\begin{equation}\label{res}
\omega-k_{_{\parallel}}\upsilon_{_{\parallel}}-k_xu_x = 0,
\end{equation}
where by $k_{_{\parallel}}$ and $\upsilon_{_{\parallel}}$ we denote
the wave vector's and velocity's longitudinal (along the magnetic
field line) components respectively and $k_x$ is the wave vector's
component along the drift. According to the theory developed by
\cite{shapo1} the plasma component with the Lorentz factor
$\gamma_p$ and the beam component with $\gamma_b$ lead to the
unstable Cherenkov drift waves with the increment given by
\begin{equation}\label{incr}
\Gamma =
\frac{\pi}{2}\frac{\omega_b^2}{\omega}\frac{\gamma_b}{\gamma_p^2},
\end{equation}
where by
\begin{equation}\label{om}
\omega = \frac{\omega_b\gamma_bc}{\gamma_p^{3/2}u_x}
\end{equation}
we denote the frequency of Cherenkov emission, $\omega_b =
\sqrt{4\pi n_be^2/m}$ is the plasma frequency of the beam component
and $n_b$ is the corresponding number density \citep{oc13}.

For considering the QLD one has to note that it is governed by two
forces. One is the synchrotron radiation reaction force with the
following components
\begin{equation}\label{f}
    F_{\perp}\approx -\alpha\frac{p_{\perp}^3}{p_{\parallel}m^{2}c^{2}},\;\;\;F_{\parallel}\approx -\frac{\alpha}{m^{2}c^{2}}p_{\perp}^{2},
\end{equation}
where $p_{\perp}$ and $p_{\parallel}$ are transversal and
longitudinal components of momentum respectively and
$\alpha=2e^{2}\omega_{B}^{2}/3c^{2}$. On the other hand, the
relativistic particles moving in a non uniform magnetic field
experience another force responsible for the conservation of the
adiabatic invariant $I = 3cp_{\perp}^2/2eB$. The corresponding
components of this force are given by \citep{landau}
\begin{equation}\label{hper}
H_{\perp} = -\frac{c}{\rho}p_{\perp},\;\;\;\; H_{\parallel} =
\frac{c}{\rho p_{\parallel}}p_{\perp}^2.
\end{equation}
It can be shown that for relatively small pitch angles when the
following condition $\partial/\partial p_{\perp}>>\partial/\partial
p_{\parallel}$ is satisfied, the kinetic equation governing the QLD
writes as \citep{machus1,malmach,oc13}
\begin{eqnarray} \label{kin2}
    \frac{\partial\textit{f }^{0}}{\partial
    t}+\frac{1}{p_{\perp}}\frac{\partial}{\partial
    p_{\perp}}\left(p_{\perp}\left[
    F_{\perp}+H_{\perp}\right]\textit{f }^{0}\right)
    =\frac{1}{p_{\perp}}\frac{\partial}{\partial p_{\perp}}\left(p_{\perp}
D_{\perp,\perp}\frac{\partial\textit{f }^{0}}{\partial
p_{\perp}}\right).
\end{eqnarray}
where by $\textit{f }^{0}\left(\mathbf{p}\right)$ we denote the
distribution function of electrons,
\begin{equation}\label{dkoef}
D_{\perp,\perp} =
8\pi\left(\frac{u_x}{c}\right)^8\left(\frac{e}{mc}\right)^2\frac{W}{\Gamma_k}
\end{equation}
is the transversal coefficient of diffusion and $W =
\gamma_bn_bmc^2$ \citep{oc13}.

\begin{figure}
  \resizebox{\hsize}{!}{\includegraphics[angle=0]{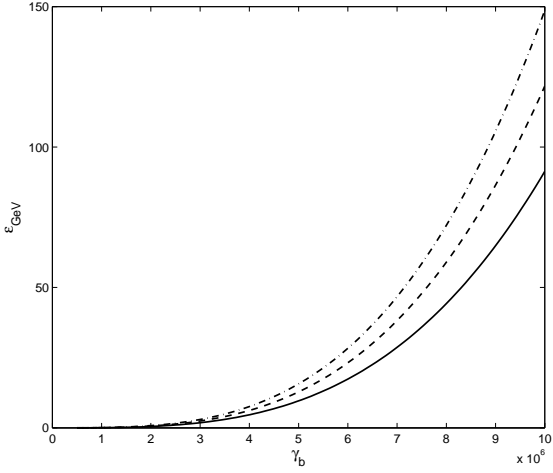}}
  \caption{Dependence of energy of synchrotron photons on the
  Lorentz factors of the beam electrons for
different values of plasma component's Lorentz factors: $\gamma_p =
1$ (solid line), $\gamma_p = 10$ (dashed line), $\gamma_p = 50$
(dashed-dotted line). The set of the parameters is: $P = P_{_{Cr}}$,
$\dot{P}=\dot{P}_{_{Cr}}$, $n_b = n_{_{GJ}}$, $R = 10^6$cm.
$\gamma_b$ varies in the interval $[0.5; 1]\times
10^7$.}\label{fig1}
\end{figure}

By combining Eqs. (\ref{f},\ref{hper}) and by taking into account
that the pitch angle $\psi$ is given by $p_{\perp}/p_{\parallel}$
one can derive for the typical parameters of Crab-like pulsars the
following relation between the corresponding components of the
forces
$$\frac{H_{\perp}}{F_{\perp}}\approx 1.1\times 10^{-4}\times
    \frac{P_{_{Cr}}}{P}\times
    \frac{\dot{P}_{_{Cr}}}{\dot{P}}\times$$
\begin{eqnarray} \label{ratio}
\;\;\;\;\;\;\;\;\;\;\;\;\;\;\;\;\;\;\;\;\;\;\;\;\times\frac{R_{lc}}{\rho}
\times\frac{10^7}{\gamma_{b}}\times\left(\frac{0.01
rad}{\psi}\right)^{2},
\end{eqnarray}
where $P$ is the pulsar's rotation period, $\dot{P}\equiv |dP/dt|$
is the modulus of its time derivative, $\dot{P}_{_{Cr}}\approx
4.21\times 10^{-13}$ss$^{-1}$ is the corresponding value for the
Crab pulsar and we have taken into account that the magnetic field
has a dipolar behavior and close to the pulsar's surface its
induction is given by $B\approx 3.2\times
10^{19}\times\sqrt{P\dot{P}}$G \citep{GJ}. From the above estimation
it is clear that $H_{\perp}$ can be neglected for the realistic
physical parameters. Therefore Eq. (\ref{ratio}) for the stationary
regime $\left(\partial/\partial t = 0\right)$ reduces to the
equation
\begin{eqnarray} \label{kin23}
    \frac{\partial}{\partial
    p_{\perp}}\left(p_{\perp}F_{\perp}\textit{f }^{0}\right)
    =\frac{\partial}{\partial p_{\perp}}\left(p_{\perp}
D_{\perp,\perp}\frac{\partial\textit{f }^{0}}{\partial
p_{\perp}}\right),
\end{eqnarray}
with the following solution \citep{oc13}
\begin{equation}\label{ff}
    \textit{f}(p_{\perp}) = Ce^{-\left(\frac{p_{\perp}}{p_{\perp_{0}}}
    \right)^{4}},
\end{equation}
where
\begin{equation}\label{pp0}
     p_{\perp_{0}}=\left(\frac{4\gamma_bm^3c^3D_{\perp,\perp}}{\alpha}\right)^{1/4}.
\end{equation}
Since the particles are distributed with the transverse momentum, it
is worthwhile to define the average value of the pitch angles
\begin{equation}\label{pitch}
\langle \psi\rangle
 = \frac{1}{p_{\parallel}}\frac{\int_{0}^{\infty}p_{\perp} \textit{f}(p_{\perp})dp_{\perp}}
 {\int_{0}^{\infty}\textit{f}(p_{\perp})dp_{\perp}}
= \frac{\sqrt{\pi}}{4\Gamma
(\frac{5}{4})}\frac{p_{\perp_{0}}}{p_{\parallel}},
\end{equation}
leading to the following energy of synchrotron photons
\citep{Lightman}
\begin{equation}\label{eps}
\epsilon_{_{GeV}}\approx 3\times 10^{-18}\gamma_b^2
B_{lc}\frac{\sqrt{\pi}}{\Gamma (\frac{5}{4})}
 \frac{p_{\perp_{0}}}{p_{\parallel}}.
\end{equation}
where $\Gamma(x)$ is the gamma function.

\section{Main results}
In this section we apply the theoretical model to the Crab-like
pulsars. Since the QLD is a result of the ChDI, it is important to
know how efficient is the corresponding process. By considering the
so-called beam and plasma components, one obtains from Eq.
(\ref{incr})
\begin{equation}\label{incr2}
\Gamma\approx 1.4\times
10^4\times\gamma_p^{-1/2}\times\frac{\gamma_b}{10^7}\times\frac{R_{lc}}{\rho}
\times\left(\frac{P_{_{Cr}}}{P}\right)^{3/4}\times\left(\frac{\dot{P}_{_{Cr}}}{\dot{P}}\right)^{1/4}
\end{equation}
It is clear that for the typical parameters of the Crab-like pulsars
the instability timescale $t_{ins}\sim 1/\Gamma\sim 10^{-4}$s, which
is much less than the kinematic timescale, which indicates the high
efficiency of the ChDI, generating the relatively low energy photons
with energies (see Eq. (\ref{om}))
\begin{equation}
\label{lowen} \epsilon_{keV}^{_{Ch}}\approx
4.7\times\gamma_p^{-3/2}\times\frac{\rho}{R_{lc}}\times\left(\frac{P}
{P_{_{Cr}}}\right)^{1/4}\times\left(\frac{\dot{P}}{\dot{P}_{_{Cr}}}\right)^{3/4}.
\end{equation}

As we see, in the magnetospheres of Crab-like pulsars the Cherenkov
drift resonance might lead to the hard $X$-rays. In the previous
section we have seen that this emission inevitably leads to the non
vanishing pitch angles (see Eq. (\ref{eps})), generating the
synchrotron emission. In Fig. \ref{fig1} we show the behavior of
energy of synchrotron photons, $\epsilon_{_{GeV}}$ versus $\gamma_b$
in the interval $[0.5; 1]\times 10^7$ for different values of
Lorentz factors of the plasma component: $\gamma_p = 1$ (solid
line), $\gamma_p = 10$ (dashed line), $\gamma_p = 50$ (dashed-dotted
line). We have taken into account the dipolar character of the
magnetic field, $B\approx 3.2\times
10^{19}\times\sqrt{P\dot{P}}\times R^3/R_{lc}^3$G \citep{GJ}, where
$R = 10^6$cm is the neutron star's radius. The rest of the
parameters is: $P = P_{_{Cr}}$, $\dot{P}=\dot{P}_{_{Cr}}$, $n_b =
n_{_{GJ}}$, where $n_{_{GJ}} = B/Pec$ is the Goldreich-Julian number
density \citep{GJ}. As it is evident from the plots by increasing
$\gamma_p$, the resulting emission energy increases as well, which
directly follows from the analytical behaviour,
$\epsilon_{_{GeV}}\propto\gamma_p^{1/8}$ (this can be seen by
combining Eqs. (\ref{dkoef},\ref{pp0}-\ref{eps})). We see that the
QLD by means of the feedback of ChDI guarantees the synchrotron
emission over $100$GeV. We can roughly estimate the possible VHE
luminosity by multiplying the number of particles involved in the
process with the single emission power, $P_{syn}\approx
2e^4\gamma_b^2B^2\langle \psi\rangle^2/3m^2c^3$. It is worth noting
that the thickness of the layer, $d$, where the electrons accelerate
to the highest energies, is of the order of $R_{lc}/2\gamma_b$.
Then, by taking into account the corresponding number of particles,
$4\pi R_{lc}^2n_{_{GJ}}d$, one can show that in the emission energy
band $150$GeV the highest energy electrons lead to the bulk
luminosity of the order of $10^{33}$erg s$^{-1}$ which is in a good
agreement with observations \citep{lessard,albert}.

\begin{figure}
  \resizebox{\hsize}{!}{\includegraphics[angle=0]{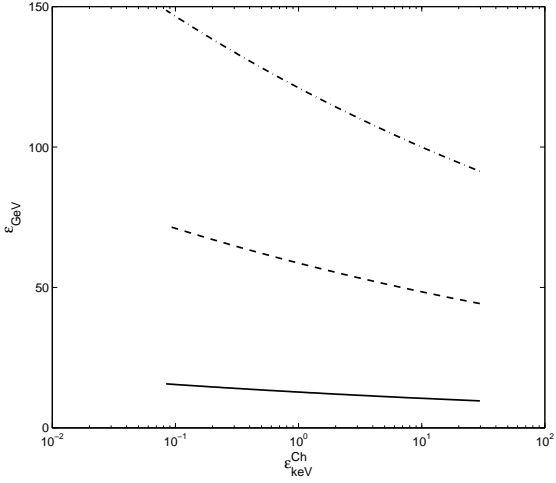}}
  \caption{Behaviour of $\epsilon_{_{GeV}}(\epsilon_{keV}^{_{Ch}})$ for
different values of the Lorentz factors of the beam component:
$\gamma_b = 5\times 10^6$ (solid line), $\gamma_b = 8\times 10^6$
(dashed line), $\gamma_b = 10^7$ (dashed-dotted line). The set of
the parameters is: $P = P_{_{Cr}}$, $\dot{P}=\dot{P}_{_{Cr}}$, $n_b
= n_{_{GJ}}$, $R = 10^6$cm. $\gamma_p$ varies in the interval $[1;
50]$.}\label{fig2}
\end{figure}

An interesting feature of this mechanism is that the VHE radiation
is strongly linked to the lower energy emission provided by the
ChDI. In Fig. \ref{fig2} we show plots for
$\epsilon_{_{GeV}}(\epsilon_{keV}^{_{Ch}})$ for three values of the
beam Lorentz factors $\gamma_b = 5\times 10^6$ (solid line),
$\gamma_b = 8\times 10^6$ (dashed line), $\gamma_b = 10^7$
(dashed-dotted line). The set of the parameters is: $P = P_{_{Cr}}$,
$\dot{P}=\dot{P}_{_{Cr}}$, $n_b = n_{_{GJ}}$, $R = 10^6$cm and the
plasma component lorentz factor varies in the interval $[1; 50]$. We
see that for the mentioned physical parameters the Cherenkov drift
photons have energies in the interval $\sim[0.02; 30]$keV, which are
strongly correlated with the VHE emission in the energy band $[20;
150]$GeV.

\begin{figure}
  \resizebox{\hsize}{!}{\includegraphics[angle=0]{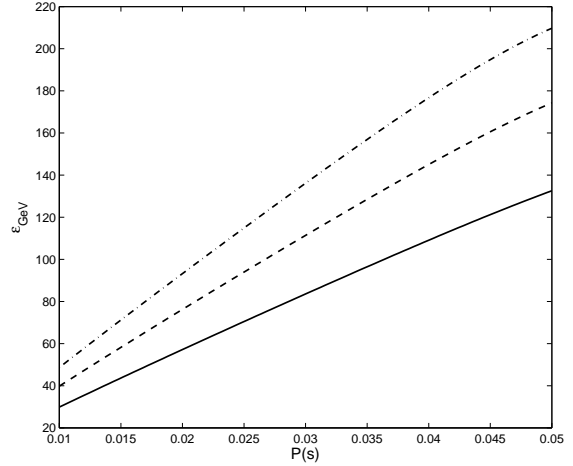}}
  \caption{Dependence of $\epsilon_{_{GeV}}$ versus $P$ for three
different values of $\gamma_p$: $\gamma_p = 1$ (solid line),
$\gamma_p = 10$ (dashed line), $\gamma_p = 50$ (dashed-dotted line).
The set of the parameters is: $P = [0.01; 0.05]s$, $\gamma_b =
10^7$, $\dot{P}=\dot{P}_{_{Cr}}$, $n_b = n_{_{GJ}}$, $R =
10^6$cm.}\label{fig3}
\end{figure}

It is interesting to study the generation of VHE synchrotron
emission for a variety of Crab-like pulsars. For this purpose in
Fig. \ref{fig3} we show the dependence of energy of synchrotron
photons on pulsar's period for different values of $\gamma_p$:
$\gamma_p = 1$ (solid line), $\gamma_p = 10$ (dashed line),
$\gamma_p = 50$ (dashed-dotted line). The set of the parameters is:
$P = [0.01; 0.05]s$, $\gamma_b = 10^7$, $\dot{P}=\dot{P}_{_{Cr}}$,
$n_b = n_{_{GJ}}$, $R = 10^6$cm. From the figure it is clear that
depending on the period, for the above parameters, the synchrotron
emission might provide energies up to $\sim 200$GeV.

The investigation shows that the Crab-like pulsars might provide VHE
synchrotron radiation strongly connected with the Cherenkov drift
emission.

\section{Summary}

The main aspects of the present work can be summarized as follows:
\begin{enumerate}

      \item We have studied the VHE synchrotron emission process driven by the feedback
      of the ChDI on the distribution function of electrons in the Crab-like
      pulsars.

      \item Considering the two component magnetospheric plasmas, we have
      developed the analytical model for studying the QLD. For this
      purpose the kinetic equation governing the mentioned process has been
      applied and analysed for the light cylinder zone. It has been
      found that under certain conditions the ChDI is very
      efficient, resulting in the creation of
      the pitch angles, which in turn prevents the synchrotron
      mechanism from damping despite the strong magnetic field.

      \item Analyzing the radiation mechanism versus several
      physical parameters, we have shown that the Crab-like pulsars
      might provide synchrotron photons with $100$s of GeV strongly
      correlated with the Cherenkov drift emission from the extreme
      UV ($20$eV) to the hard $X$-rays ($30$keV).

      \end{enumerate}

\section*{Acknowledgments}

The research was partially supported by the Shota Rustaveli National
Science Foundation grant (N31/49).

\bsp

\label{lastpage}


\begin{thebibliography}{99}
\bibitem[Albert et al.(2008)]{albert} Albert J., Aliu, E., Anderhub, H., et al., 2008, ApJ, 674, 1037
\bibitem[Aleksi\'c et al.(2011)]{magic1} Aleksi\'c, J., Alvarez, E. A., Antonelli, L. A., et al. 2011, ApJ, 744, 43
\bibitem[Aleksi\'c et al.(2012)]{magic2} Aleksi\'c, J., Alvarez, E. A., Antonelli, L. A., et al. 2012, A\&A, 540, 69
\bibitem[Aliu et al.(2011)]{veritas} Aliu, E., Arlen, T., Aune, T., et al. 2011, Sci, 334, 69
\bibitem[Chkheidze et al.(2013)]{cmo13} Chkheidze, N., Machabeli, G. \& Osmanov, Z., 2013, ApJ,
773, 7
\bibitem[Chkheidze et al.(2011)]{cmo11} Chkheidze, N.,
Machabeli, G. \& Osmanov, Z., 2011, ApJ, 730, 62
\bibitem[Chkheidze \& Osmanov(2012)]{co12} Chkheidze, N. \& Osmanov, Z., 2012, MNRAS,
419, 239
\bibitem[Goldreich \& Julian(1969)]{GJ} Goldreich, P. \& Julian, W. H., 1969, ApJ,
157, 869
\bibitem[Kazbegi et al.(1991)]{kmm} Kazbegi, A. Z.,
Machabeli, G. Z., \& Melikidze, G. I. 1991, MNRAS, 253,
377
\bibitem[Landau \& Lifshitz(1971)]{landau} Landau, L. D., \&
Lifshitz, E. M. 1971, Classical Theory of Fields (London:
Pergamon)
\bibitem[Lessard et al.(2000)]{lessard} Lessard, R. W.,
Bond, I. H., Bradbury, S. M., et al., 2000, ApJ, 531, 942
\bibitem[Lominadze et al.(1979)]{lomin} Lominadze J.G.,
Machabeli G.Z. \& Mikhailovsky A.B., 1979, J. Phys. Colloq., 40, No.
C-7, 713
\bibitem[Machabeli \& Osmanov(2010)]{mo10} Machabeli, G. \&
Osmanov Z., 2010, ApJ, 709, 547
\bibitem[Machabeli \& Osmanov(2009)]{mo09} Machabeli, G. \& Osmanov Z., 2009, ApJL, 700, 114
\bibitem[Machabeli \& Usov(1979)]{machus1} Machabeli G.Z. \& Usov V.V., 1979, AZhh Pis'ma, 5, 445
\bibitem[Mahajan et
al.(2013)]{screp1} Mahajan, S., Machabeli, G., Osmanov, Z. \&
Chkheidze, N., 2013, Nat. Sci. Rep. 3, 1262
\bibitem[Malov \& Machabeli(2001)]{malmach} Malov I.F. \&
Machabeli G.Z., 2001, ApJ, 554, 587
\bibitem[Osmanov(2011)]{osm11}
Osmanov Z., 2011, MNRAS, 411, 973
\bibitem[Osmanov \& Chkheidze(2013)]{oc13} Osmanov Z. \& Chkheidze, N., 2013, ApJ, 764, 59
\bibitem[Osmanov et al.(2015)]{screp2} Osmanov, Z., Mahajan, S., Machabeli, G. \& Chkheidze, N., 2015,
Nat. Sci. Rep., 5, 14443
\bibitem[Osmanov Z. \& Rieger(2009)]{or09} Osmanov Z. \& Rieger F., 2009, A\&A, 502, 15
\bibitem[Rybicki \& Lightman(1979)]{Lightman} Rybicki,  G.B. \& Lightman, A. P., 1979,
Radiative Processes in Astrophysics. Wiley, New
York
\bibitem[Shapakidze et al.(2003)]{shapo2} Shapakidze, D.,
Machabeli, G., Melikidze, G. \& Khechinashvili, D., 2003, Phys. Rev.
E, 2, 026407
\bibitem[Shapakidze et al.(2002)]{shapo1} Shapakidze, D., Melikidze, G.
\& Machabeli, G., 2002, A\&A, 394, 729
\bibitem[Sturrock(1971)]{stur}
Sturrock, P. A., 1971, ApJ, 164, 529
\bibitem[Tademaru(1973)]{tadem} Tademaru, E., 1973, ApJ, 183, 625



\end{thebibliography}
\end{document}